\documentclass[a4paper,fleqn,10pt]{cas-dc}
\usepackage{algorithm}
\usepackage{algorithmic}
\usepackage{diagbox}
\usepackage[authoryear]{natbib}
\setcitestyle{comma}
\usepackage{xurl}
\usepackage{enumitem} 
\usepackage{hyperref}
\usepackage{subcaption}
\usepackage{xcolor,colortbl}
\definecolor{novelartifact}{RGB}{255,200,200}
\usepackage{soul}
\usepackage{balance}

 \usepackage{float}    
\usepackage{caption}

\renewcommand{\paragraph}[1]{\vspace{0.03in}\noindent{\bf{#1}.}}
\begin{document}
\let\WriteBookmarks\relax
\let\printorcid\relax
\def\floatpagepagefraction{1}
\def\textpagefraction{.001}    
\title [mode = title]{Memory Forensics Techniques for Automated Detection and Analysis of Go Malware}
\shorttitle{}
\author[1]{Hala Ali}
\cormark[1]
\ead{alih16@vcu.edu}
\affiliation[1]{organization={Department of Computer Science},
                addressline={Virginia Commonwealth University}, 
                country={USA}}

\author[2]{Andrew Case}
\ead{andrew@dfir.org}
\affiliation[2]{organization={Volatility Foundation},
                country={USA}}
\author[1]{Irfan Ahmed}
\ead{iahmed3@vcu.edu}

\cortext[cor1]{Corresponding author}

\begin{abstract}
The Go programming language has become increasingly popular among malware developers due to its ability to produce statically linked, cross-platform executables that challenge traditional analysis techniques. These binaries embed a substantial runtime and compiler-generated metadata and are compiled with aggressive optimizations that discard type information for function parameters and local variables. Go's design further complicates analysis by representing strings as pointer–length pairs rather than null-terminated sequences, employing a caller-allocated stack model that obscures argument boundaries, and fragmenting program state across concurrent goroutines. Although existing static analysis and reverse engineering tools provide Go-specific support, they remain limited to compile-time artifacts and cannot recover runtime execution state and artifacts that persist solely in memory. To address this gap, we present the first memory forensics framework for runtime analysis of Go binaries. By parsing Go's internal structures, our framework reconstructs type and function metadata, recovers heap-allocated and static strings, and distinguishes application-level functions. Through ABI-aware backward analysis, it derives execution paths and argument values from call sites. To capture runtime state beyond what static analysis reveals, it analyzes goroutine stacks to identify actively executing functions and recover their runtime argument values. We implemented all capabilities as Volatility 3 plugins and evaluated them against malware observed in recent incidents, such as the \textit{BRICKSTORM} backdoor, \textit{Obscura} ransomware, and \textit{Pantegana} RAT, as well as open-source samples for reproducibility. The framework successfully recovered C2 endpoints, persistence mechanisms, encryption keys, ransom notes, and execution state, including critical runtime artifacts that were absent from published threat intelligence.
\end{abstract}

\begin{keywords}
Go \newline Golang  \newline Memory Forensics \newline Malware Analysis  \newline Volatility 3
\end{keywords}
\maketitle
\vspace{-0.1in}
\section{Introduction}
\label{s1}

The \textit{Go} programming language has become increasingly popular among malware authors due to its cross-platform compilation capabilities, which allow a single codebase to produce executables targeting multiple platforms \citep{unit42_golang, maurya_crowdstrike, stevens_bitsight}. This adoption has created significant challenges for malware analysts and incident responders. Go produces statically linked binaries that embed a substantial custom runtime and extensive compiler-generated metadata, resulting in large executables \citep{d4rksystem}. Moreover, such binaries are compiled with aggressive optimizations that discard type information for function parameters and local variables \citep{instatunnel2025rustgo}. These properties reduce the effectiveness of traditional detection mechanisms and complicate reverse engineering.

Go's design introduces further obstacles for analysis. Unlike C and C++, strings are represented as pointer–length pairs rather than null-terminated byte sequences, eliminating explicit delimiters and thus making reliable extraction of string-based artifacts difficult \citep{alphagolang_2021}. Dynamic analysis is similarly challenging, as Go employs a caller-allocated stack model in which arguments and return values reside in caller-reserved memory regions, obscuring argument boundaries and complicating parameter recovery \citep{pnfsoftware}. These inherent challenges are compounded when binaries are stripped of symbols, lack debug information, or are obfuscated using tools such as Garble, which hashes identifiers and removes metadata to disrupt static analysis \citep{raimbaud_5}.

Although reverse-engineering tools, including \textit{IDA Pro} and \textit{Ghidra}, have improved their support for Go binaries \citep{hexrays_ida_go_2025, trellix_feeding_gophers_2023}, they rely on heuristic-based disassembly without fully parsing the embedded metadata. This results in incomplete function boundaries, fragmented call graphs, and prototypes lacking type annotations. Specialized tools, such as \textit{GoReSym}, address these limitations by parsing metadata structures derived from the Go runtime source code \citep{mandiant_goresym_github, goresym_google}. Nevertheless, all existing tools remain limited to compile-time artifacts and cannot reconstruct runtime program state or recover critical runtime artifacts that reside solely in memory, such as decrypted strings, encryption keys, and command-and-control configurations.

To address this gap, this paper introduces the first memory forensics framework for runtime analysis of Go binaries. By parsing Go's internal structures, including \texttt{pclntab} and \texttt{moduledata}, it recovers function metadata, type descriptors, and interface tables. Leveraging this metadata, the framework identifies heap objects, extracts dynamically allocated and static strings, and classifies functions by origin. Through ABI-aware backward analysis, it derives execution paths and argument values from call sites. Goroutine stack analysis extends this to runtime state, identifying actively executing functions and recovering argument values  that exist only at execution time. By relying on structural validation rather than symbolic information, the framework remains effective on stripped binaries and maintains robustness against obfuscation that corrupts version identifiers and magic bytes within \texttt{pclntab}.

We implemented our framework as a suite of Volatility 3 plugins \citep{volatility_framework}, supporting multiple Go versions on both Linux and Windows.  We evaluated  it against real-world malware recently observed in the wild, including the \textit{BRICKSTORM} backdoor \citep{cisa_brickstorm_2025,mandiant_brickstorm_espionage_2025}, \textit{Pantegana} RAT \citep{huntio_pantegana_rat, huntio_ghost_pantegana_2025}, \textit{Obscura} ransomware \citep{huntress_obscura_2025, coveware_obscura_dataloss_2025}, as well as the open-source \textit{Screenshotter} application for reproducibility \citep{omaidf_go_malware_github}. 
Results demonstrate that \texttt{go\_strings} successfully extracted string artifacts with clear boundaries and memory locations, such as hardcoded credentials from \textit{Screenshotter}, which appear only as concatenated data in standard \texttt{strings} output. From \textit{Obscura}, it recovered the ransom note, Curve25519 public key, and threat actor communication channels, including a \textit{Tor} hidden service URL and \textit{TOX} messaging ID. The \texttt{go\_functions} plugin recovered complete binary paths and classified functions by source file, while its ABI-aware backward analysis of \textit{BRICKSTORM} revealed C2 endpoints, persistence paths, and DNS-over-HTTPS resolver settings. Finally, \texttt{go\_goroutines} analyzed \textit{Pantegana} goroutine stacks, recovering deployment-specific C2 configuration from active function arguments, such as  server address, API endpoints, and beacon intervals.
We summarize the contributions of this paper as follows:

\begin{itemize}[noitemsep, nolistsep, leftmargin=0.18in]
   \item We present the first memory forensics framework for reconstructing the runtime behavior of Go binaries, enabling analysis beyond compile-time artifacts.
    
    \item We parse Go's internal structures to extract strings from heap and static sections, classify functions by origin, derive execution paths and argument values, and analyze goroutine stack to recover execution state.
    
    \item We implement the proposed framework as Volatility 3 plugins and evaluate it on real-world Go malware, demonstrating accurate recovery of critical runtime forensic artifacts.
\end{itemize}

The paper is organized as follows. Section \ref{s2} summarizes related work. Section \ref{s3} explains the framework. Section \ref{s4} presents the evaluation, Section \ref{s5} discusses limitations and future directions, and Section \ref{s6} concludes the paper.
\vspace{-0.1in}
\section{Related Work}
\label{s2} 
Prior work has shown that malware frequently abuses userland runtimes, motivating memory forensics techniques that reconstruct high-level execution state from process memory. The Android runtime, for example, has been widely targeted to access sensitive resources, such as call history, text messages, microphones, and cameras \citep{smmarwar2024android}. In response, multiple memory analysis approaches have been proposed to acquire Android memory and detect runtime-resident malicious behavior \citep{sylve2012acquisition, case2017memory, ali2020app, ali2019droidscraper, tam2015detecting}. On macOS, the Objective-C and Swift runtimes have similarly been analyzed using Volatility plugins to detect suspicious method invocation, keystroke logging, and runtime manipulation \citep{case2016detecting}. Subsequent research extended these capabilities to enumerate loaded classes and instances, decode type information, and identify suspicious API usage and malicious function calls \citep{manna2021modern}. The .NET and .NET Core runtimes have also been subject to analysis that extract memory-resident assemblies and recover runtime components, such as classes, fields, methods, and native imports \citep{manna2022memory}. More recently, researchers analyzed the V8 JavaScript runtime by extracting and tracing V8 objects via internal structures such as \textit{MetaMap} \citep{wang2022juicing}, and analyzed the Python runtime by recovering modules, classes, functions, and execution context to capture malicious behavior \citep{ali2025memory, ali2025leveraging}.
\vspace{-0.1in}
\section{Memory Analysis Framework}
\label{s3}
This section presents our memory forensics framework, which recovers Go’s embedded runtime metadata as the foundation for all subsequent analysis.

\vspace{-0.05in}
\subsection{Go Runtime Metadata Extraction}
\label{s3-1}
Go binaries embed linker-generated metadata that supports core runtime functionality, including stack unwinding and garbage collection. A primary component of this metadata is \texttt{pclntab} (program counter line table), which begins with a \texttt{pcHeader} structure. The \texttt{pcHeader} contains version-identifying magic bytes, architectural parameters, such as the minimum instruction size (\texttt{minLC}) and pointer size (\texttt{ptrSize}), counts of functions and source files (\texttt{nfunc}, \texttt{nfiles}), and relative offsets to several internal tables. These tables include \texttt{funcnametab}, which stores function name strings; \texttt{filetab}, which stores source file paths; \texttt{cutab}, which maps compilation units to ranges of file indices; \texttt{pctab}, which encodes PC-value data used for stack frame size and line number resolution; and \texttt{pclntable}, which contains the function table (\texttt{ftab}) and associated \texttt{\_func} structures. Each \texttt{ftab} entry maps a function’s entry program counter to its corresponding \texttt{\_func} structure, which  references the function name, PC-value streams, and source file information.

\begin{figure*}
    \centering
    \includegraphics[width=0.85\textwidth]{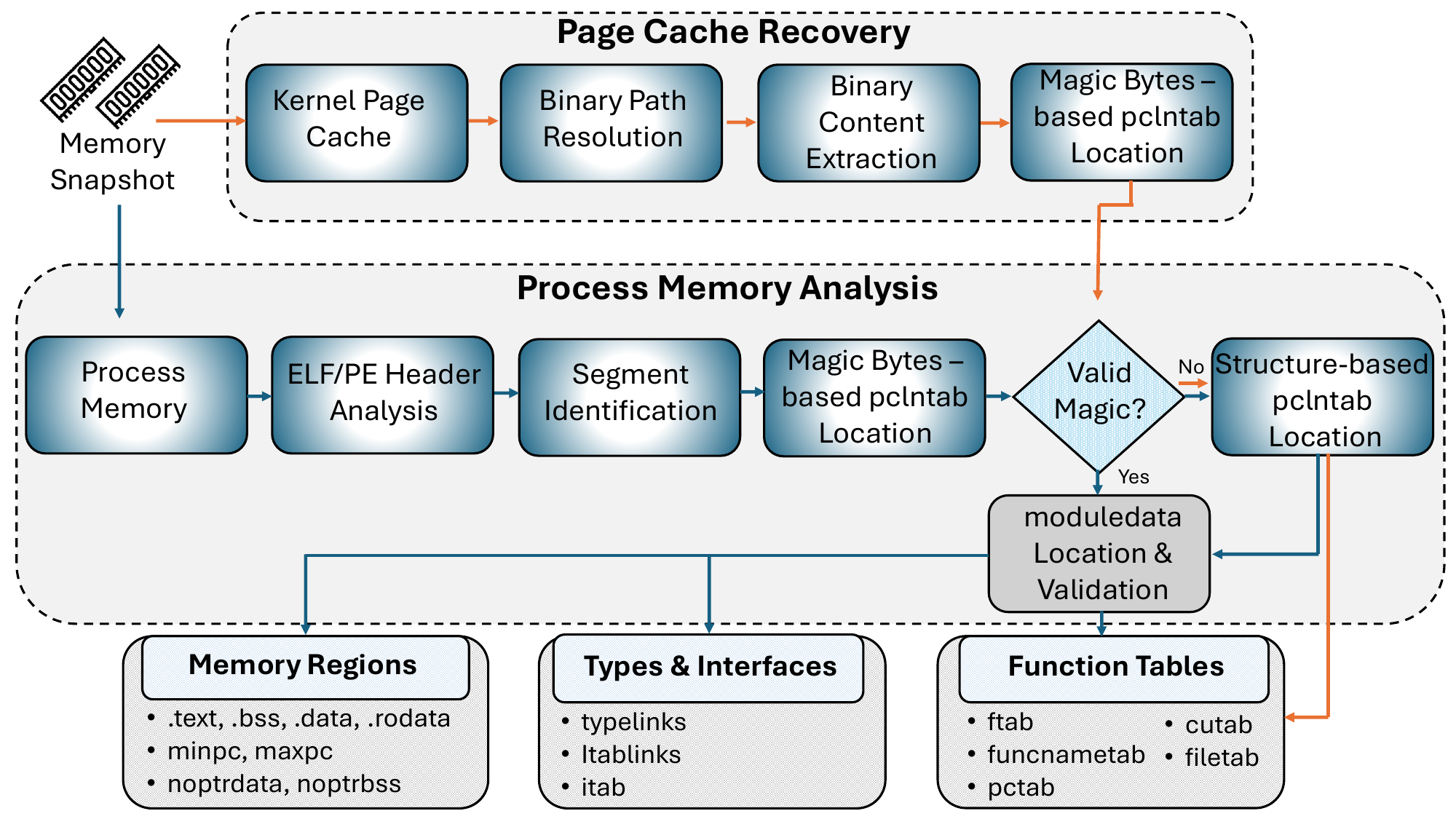}
    \vspace{-0.05in}
    \caption{Go runtime metadata extraction}
    \label{extraction_pipeline}
    \vspace{-0.2in}
\end{figure*}

When a Go binary is loaded, the runtime initializes \texttt{moduledata}, a linker-emitted structure located in a writable data section. This structure references the same internal tables as \texttt{pcHeader}, but uses absolute addresses resolved at process initialization rather than relative offsets. In addition, \texttt{moduledata} defines memory region boundaries for the loaded binary, including \texttt{text}, \texttt{data}, \texttt{bss}, \texttt{rodata}, and \texttt{types}. It also contains two slice fields: \texttt{typelinks}, which store offsets to type descriptors, and \texttt{itablinks}, which store pointers to interface tables (\texttt{itab}) that bind concrete types with implemented interfaces. Figure~\ref{extraction_pipeline} summarizes the metadata extraction pipeline. The primary path recovers metadata from process memory, while a supplementary path extracts paged-out function names and source file paths from the cached binary.

\vspace{-0.05in}
\subsubsection{Process Memory Analysis}
\begin{sloppypar}
The framework parses the executable header to identify program segments and then scans read-only segments for \texttt{pclntab} using version-specific magic bytes: \verb|\xf1\xff\xff\xff| (Go 1.20+), \verb|\xf0\xff\xff\xff| (Go 1.18--1.19), \verb|\xfa\xff\xff\xff| (Go 1.16--1.17), and \verb|\xfb\xff\xff\xff| (Go 1.2--1.15). Each candidate match is validated by enforcing \texttt{pcHeader} invariants, requiring zeroed padding bytes, \texttt{minLC} $\in \{1,2,4\}$, \texttt{ptrSize} $\in \{4,8\}$, and function and file counts within reasonable bounds. When magic-byte detection fails due to obfuscation, the framework transitions to structure-based discovery, evaluating aligned addresses using these same invariants. 
Once \texttt{pclntab} is validated, the framework locates \texttt{moduledata} by scanning writable segments for a pointer equal to the recovered \texttt{pcHeader} address, as the first field of \texttt{moduledata} references \texttt{pcHeader}. Candidate structures are then validated by enforcing ordered memory regions (e.g., $\texttt{data} < \texttt{edata}$) and well-formed slice headers satisfying $\texttt{len} \leq \texttt{cap}$.

\end{sloppypar}

\vspace{-0.05in}
\subsubsection{Page Cache Recovery}
Function names and source file paths are primarily used for diagnostics (i.e., panic messages and stack traces) and are rarely accessed during normal execution, as the runtime operates directly on program counters. Consequently, pages containing \texttt{funcnametab} entries may be paged out under memory pressure, particularly for infrequently executed code. To recover complete metadata, the framework additionally extracts \texttt{pclntab} from the kernel page cache. In ELF binaries, \texttt{pclntab} resides in the \texttt{.gopclntab} section, and, in PE binaries, it is embedded as the \texttt{runtime.symtab} symbol within the \texttt{.symtab} section. In contrast, \texttt{moduledata} cannot be reconstructed from page cache alone, as it is only partially initialized in the static binary and is finalized by the runtime during process startup.


\vspace{-0.05in}
\subsubsection{Obfuscation Handling}
Go obfuscation tools, such as \textit{Garble}, explicitly target symbolic and metadata-level artifacts~\citep{mvdan_garble_changelog_2024}, including function names, file paths, type names, and static strings, to hinder static analysis and reverse engineering. These transformations do not modify the core components required by the Go runtime for correct execution. Accordingly, our framework avoids reliance on symbolic information and instead analyzes runtime structures directly, as discussed earlier in this section. Heap spans and goroutine objects are recovered based on their structural layout rather than by resolving the \texttt{runtime.mheap} or \texttt{runtime.allgs} symbols. Consequently, recovery of critical forensic artifacts, such as function bodies, argument layouts, type descriptors, heap objects, and goroutine stacks, remains robust under obfuscation. In addition, dynamically allocated strings located in the heap remain in plaintext at runtime and are therefore unaffected by compile-time obfuscation, representing a key advantage of memory forensics over static analysis.

Given that obfuscation can corrupt or randomize version identifiers and magic bytes within the \texttt{pclntab}, our framework identifies the Go version based on changes in data structures introduced across Go releases. For example, the \texttt{textStart} field in \texttt{pcHeader}, as well as the \texttt{rodata} and \texttt{gofunc} fields in \texttt{moduledata}, are introduced in Go 1.18. Coverage-related fields (\texttt{covctrs}, \texttt{ecovctrs}) and the \texttt{inittasks} field appear in Go 1.20, while the \texttt{sys.NotInHeap} marker is introduced in Go 1.21.  However, when symbolic deobfuscation is required, existing tools can be applied to executables extracted from memory. For example, \textit{GoReSym} \citep{mandiant_goresym_github} reconstructs function and package symbols using Go runtime metadata and heuristic analysis; \textit{GoStringUngarble} \citep{mandiant_gostringungarbler_github}  recovers strings obfuscated by \textit{Garble}; and \textit{GoResolver} \citep{volexity_goresolver_github} resolves obfuscated standard library functions via control-flow graph similarity. 

\vspace{-0.05in}
\subsubsection{Type Metadata Extraction}
Each entry in \texttt{typelinks} references a type descriptor beginning with a common \texttt{\_type} header, which encodes size, pointer metadata (\texttt{ptrdata}), alignment requirements (\texttt{align}, \texttt{fieldAlign}), type flags (\texttt{tflag}), and a type kind (\texttt{kind}). Go defines 26 kinds spanning primitive, composite, and abstract types \citep{pnfsoftware}. Following the common header, descriptors include kind-specific metadata sufficient to traverse and reconstruct typed values, such as pointers, slices, arrays, structs, interfaces, functions, and methods. For type methods, the descriptor is further followed by an \texttt{uncommonType} structure containing method-related metadata, including the package path, method counts, and references to method descriptors.  Each method descriptor specifies the method name, the number of input parameters and return values, the concrete types of each, and the entry points for direct (\texttt{tfn}) and interface-based (\texttt{ifn}) invocation.

\vspace{-0.05in}
\subsubsection{Interface Metadata Extraction}
Each \texttt{itab} entry stores pointers to interface type descriptor (\texttt{inter}), concrete type descriptor (\texttt{\_type}), cached type hash, and function pointer array (\texttt{fun}) used for method dispatch. The \texttt{fun} array contains one entry per interface method, ordered according to the interface’s method set, with each entry holding the program counter of the concrete implementation. We parse each \texttt{itab} by resolving \texttt{inter} to recover the interface method set and \texttt{\_type} to identify the implementing type. The \texttt{fun} entries then provide direct mappings from interface methods to concrete implementations, enabling resolution of polymorphic calls executed via interface dispatch.
\vspace{-0.05in}
\subsection{String Recovery}
\label{s3-2}
Go represents strings as a structure comprising a pointer to the underlying UTF-8 byte array and an integer specifying the length. This separates headers from backing data, with each potentially residing in different memory regions. For instance, heap-resident and stack-resident headers reference data in either the heap (for dynamically created strings) or .rodata (for static constants). In contrast, headers in static sections (.rodata, .data, .bss) always reference data located only in .rodata, as these represent compile-time literals or global variables initialized with constant strings.
\vspace{-0.05in}
\subsubsection{Heap-Based String Recovery}
\label{s3-2-1}
\begin{figure}
    \centering
    \includegraphics[width=\columnwidth]{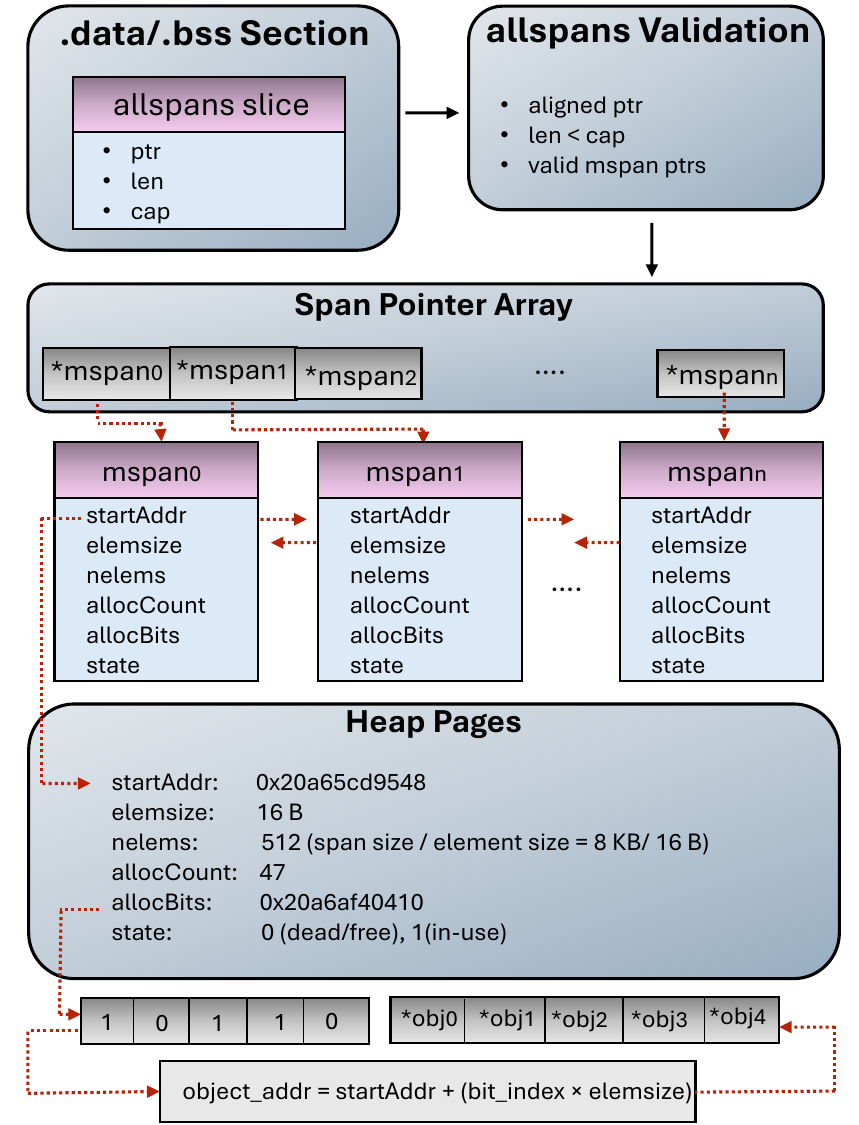}
   \vspace{-0.2in}
   \caption{Heap object recovery}
    \label{heap_strings}
\end{figure}

Recovering heap-resident strings requires identifying allocated heap objects and interpreting them using runtime type metadata. Figure~\ref{heap_strings} summarizes this process.

\paragraph{Heap Structure Parsing}
The Go heap is organized into spans containing fixed-size slots for objects of a given size class. While span recovery relies on locating the global \texttt{runtime.mheap} symbol, this structure is large, vary across versions, and may be absent in stripped or obfuscated binaries. Instead, we locate the \texttt{allspans} slice header within the \texttt{.data} or \texttt{.bss} sections and validate candidates using alignment checks, length–capacity constraints, and verification that referenced entries point to valid \texttt{mspan} structures. Each \texttt{mspan} describes a contiguous memory region divided into fixed-size slots. Key fields include the base address (\texttt{startAddr}), slot size (\texttt{elemsize}), total capacity (\texttt{nelems}), number of allocated objects (\texttt{allocCount}), allocation bitmap (\texttt{allocBits}), and span state. For active spans, allocated objects are recovered by iterating the allocation bitmap and calculating object addresses using their \texttt{bit\_index} (position in the bitmap).

\paragraph{Size-to-Type Mapping}
Heap objects lack type annotations, preventing direct identification of their types. We therefore infer candidate types by correlating the object size specified by each span’s \texttt{elemsize} with recovered type descriptors. By extracting \texttt{\_type.size} from each descriptor, we group types by size and map heap object sizes to candidate types. Since multiple types may share the same size, a single object size may correspond to several candidates. To reduce ambiguity, candidates are restricted to types whose layout may directly contain or indirectly reference string values. These include strings; arrays or slices of strings; composite types that reference string-typed elements or fields; maps whose key or value type involves strings; and interfaces, whose concrete types are determined dynamically at runtime and therefore cannot be excluded statically.

\paragraph{Type-Guided Object Interpretation}
After identifying candidate types for each object, we interpret the object under each candidate type. For objects whose size equals two machine words, we first disambiguate interface and string representations. If the first word points to a valid \texttt{itab} address, the object is treated as an interface value and its concrete payload is processed recursively. Otherwise, we treat the first word as a potential string data pointer and the second as a potential length. We then validate that the pointer references accessible memory, the length is within reasonable bounds, and the referenced bytes form valid UTF-8 with predominantly printable characters. Objects that fail string validation are excluded and evaluated under the remaining candidates. Pointer-typed objects, slices, structures, and maps (hash tables) are analyzed using the same algorithm against their contained fields.

\vspace{-0.05in}
\subsubsection{Static Section String Recovery}
Unlike the heap, static sections (\texttt{.rodata}, \texttt{.data}, and \texttt{.bss}) store data sequentially without size metadata or explicit object boundaries. To recover strings from these regions, we scan for patterns matching Go string headers. For each aligned candidate, the first word is interpreted as a potential data pointer and the second as a potential length, and candidates are validated using the same pointer, length, and content checks described earlier. This pattern-based approach enables recovery of compile-time string literals and global string variables that reference constant data.

\vspace{-0.05in}
\subsection{Function Analysis}
\label{s3-3}
\begin{figure*}[!htb]
    \centering
    \includegraphics[width=0.85\textwidth]{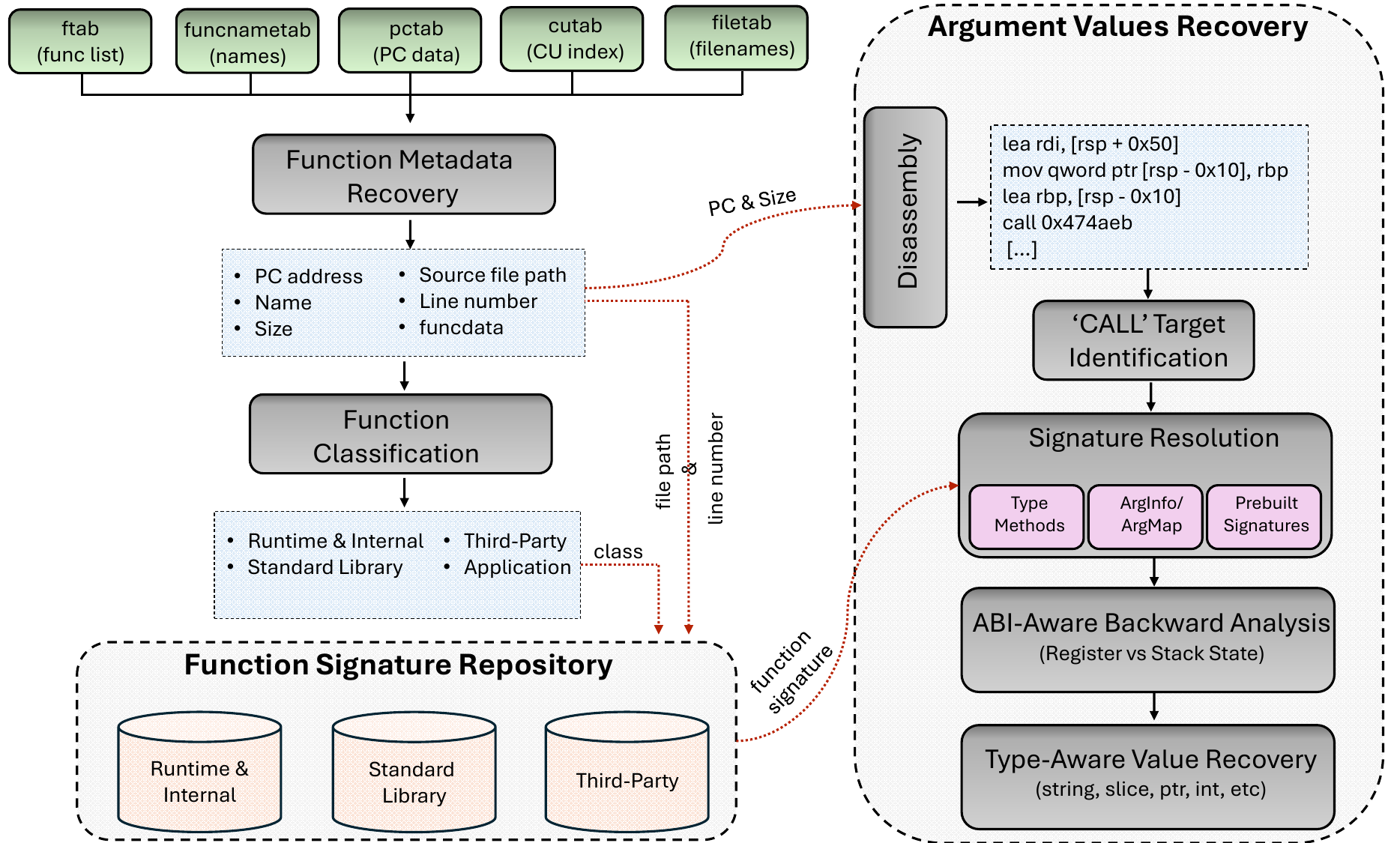}
   \vspace{-0.05in}
   \caption{Function analysis workflow}
     \vspace{-0.2in}
    \label{func_analysis}
\end{figure*}

Figure \ref{func_analysis} illustrates the function analysis workflow to recover function metadata, classify functions by origin, infer argument types, and apply ABI-aware backward analysis to recover argument values at call sites.

\vspace{-0.05in}
\subsubsection{Function Metadata Recovery}
We recover function metadata by parsing the internal tables referenced by \texttt{moduledata} (Section~\ref{s3-1}). Each function is described by a \texttt{\_func} structure, which references the function name via \texttt{nameoff} (an offset into \texttt{funcnametab}), identifies its compilation unit via \texttt{cuOffset}, and provides offsets into PC-relative metadata streams in \texttt{pctab} for stack frame sizes (\texttt{pcsp}), source file indices (\texttt{pcfile}), and line numbers (\texttt{pcln}). These streams encode metadata as compact sequences of value and program-counter deltas, allowing metadata to be resolved at arbitrary instruction addresses within the function. Although a function is defined in a single source file, inlined and compiler-generated code may originate from other files. Consequently, source file paths are not stored directly in \texttt{\_func}. Instead, the file index is decoded from the \texttt{pcfile} stream at the target program counter and combined with \texttt{cuOffset} to index into \texttt{cutab}, yielding the corresponding offset into \texttt{filetab} where the source file path resides.

\vspace{-0.05in}
\subsubsection{Function Classification}
Once the source file path is recovered, we use it to classify functions. Paths beginning with \texttt{runtime/} are classified as runtime core, while paths under \texttt{internal/} are split into runtime-critical internal packages (e.g., \texttt{internal/abi}, \texttt{internal/cpu}) and other standard-library internal packages (e.g., \texttt{internal/poll}). Paths whose first segment matches a public standard library package (e.g., \texttt{net}, \texttt{os}, \texttt{fmt}) are classified as public standard library. On the other hand, third-party functions are identified by domain-based prefixes (e.g., \texttt{github.com}, \texttt{gitlab.com}) or by paths under \texttt{vendor/} directories. Remaining functions, including those in the \texttt{main} package and other user-defined packages within the same module, are classified as application-level code.

\vspace{-0.05in}
\subsubsection{Argument Type Inference}
We leverage the recovered source file path and entry line number to resolve function identities via signature repositories, particularly when function names are unavailable due to paging. Each signature associates a function name with its defining file, entry line number, and parameter types. For type methods, signatures are recovered directly from the \texttt{uncommonType} structure. For runtime, internal, and standard library functions, we parse the official Go source distribution for the detected Go version, extracting signatures from both \texttt{.go} source files and compiler-generated \texttt{.s} assembly files. For third-party dependencies,  which vary across applications, signatures are generated on demand by parsing the corresponding source packages.

Application-level functions present a distinct challenge as their signatures are not known a priori and type information is frequently eliminated by aggressive compiler optimizations. To address this, we infer argument structure from compiler-generated metadata stored in the \texttt{funcdata} associated with each \texttt{\_func}. For binaries compiled with Go~1.17 and later, we parse \texttt{ArgInfo} bytecode, while for earlier versions we rely on \texttt{ArgsPointerMaps}. Both structures describe the layout of arguments within the call frame. \texttt{ArgInfo} encodes argument boundaries explicitly, representing scalar arguments by their frame offsets and sizes, and composite arguments (e.g., strings, slices, structs) as aggregates delimited by start and end markers.  Unlike \texttt{ArgInfo}, \texttt{ArgsPointerMaps} encodes only a bitmap indicating which stack slots within the argument area contain pointers, without explicit argument boundaries. We therefore infer boundaries using Go’s type layout conventions.

\vspace{-0.05in}
\subsubsection{ABI-Aware Backward Analysis}
Once signatures are resolved, we map parameters to their locations based on the calling convention. Go 1.17 and later use a register-based ABI, passing arguments through general-purpose registers and spilling excess arguments to the stack~\citep{go_abi_internal}. Earlier versions use a stack-based ABI, placing all arguments at consecutive word-aligned stack offsets.
To recover argument values, we disassemble the caller function and traverse backward from each \texttt{CALL} instruction to identify the last write to each argument location. Immediate \texttt{mov} instructions define constant values, register-to-register \texttt{mov} instructions propagate values through copy chains, RIP-relative \texttt{lea} instructions compute addresses of static data, and \texttt{xor reg, reg} patterns indicate zero initialization. Traversal terminates at preceding call instructions to avoid crossing call boundaries, as register contents may be modified by callees. 
\vspace{-0.05in}
\subsection{Goroutine Stack Analysis}
\label{s3-4}
Go maintains a global slice, \texttt{runtime.allgs}, containing pointers to all goroutine structures. As stripped binaries lack this symbol, we scan writable memory regions for candidate slice headers and validate them by ensuring that the referenced array contains pointers to well-formed \texttt{g} structures. A candidate \texttt{g} is considered valid if its stack bounds are properly aligned and ordered, its status field encodes a correct runtime state, and its goroutine identifier falls within reasonable bounds.

\paragraph{Stack Unwinding} 
Each goroutine's execution state is stored in the \texttt{g.sched} field, which stores the current stack pointer (\texttt{sp}) and program counter (\texttt{pc}). We reconstruct the call stack by iteratively unwinding frames starting from these values. At each iteration, the corresponding function is identified by searching the function table (\texttt{ftab}) for the entry whose PC range encompasses the current \texttt{pc}. The \texttt{pcsp} stream of the function is then decoded to obtain \texttt{sp\_delta}, representing the frame size at that \texttt{pc}. Adding \texttt{sp\_delta} to the current \texttt{sp} locates the caller's return address, which resides at the boundary between frames. Since Go's calling convention places callee arguments in caller-allocated stack space \citep{go_abi_internal}, the argument area begins immediately above the return address at \texttt{sp + sp\_delta + ptrSize}, where \texttt{ptrSize} denotes the pointer size. To advance to the caller's frame, the return address becomes the new \texttt{pc} and the argument base becomes the new \texttt{sp}. Unwinding continues until an invalid return address is encountered or the stack bounds are exceeded. Once the argument base is determined, individual argument boundaries are established using the argument type reference techniques described in Section \ref{s3-3}. 

\vspace{-0.1in}
\section{Experimental Evaluation}
\label{s4}
\begin{sloppypar}
We implemented our framework as a suite of Volatility 3 plugins, namely \texttt{go\_strings}, \texttt{go\_functions}, and \texttt{go\_goroutines}, and evaluated them using memory images collected from controlled executions of precompiled Go binaries\footnote{The plugins are available at \url{https://github.com/HalaAli198/Go-Memory-Forensics}.}. Page cache recovery leveraged existing Volatility~3 plugins, \textit{linux.pagecache} and \textit{windows.dumpfiles}, while function disassembly was performed using \textit{Capstone} \citep{capstone_engine}. Our evaluation includes Linux memory images of the \textit{BRICKSTORM} backdoor (Go 1.16.3) and \textit{Pantegana} RAT (Go 1.25.5), a Windows memory image of the \textit{Obscura} ransomware (Go 1.15), and an open-source \textit{Screenshotter} application compiled with Go 1.24.10.
\end{sloppypar}

\vspace{-0.05in}
\subsection{Screenshotter Memory Analysis}
\begin{figure*}[!htb]
  \centering
  \begin{subfigure}{0.9\textwidth}
    \centering
    \includegraphics[width=\textwidth]{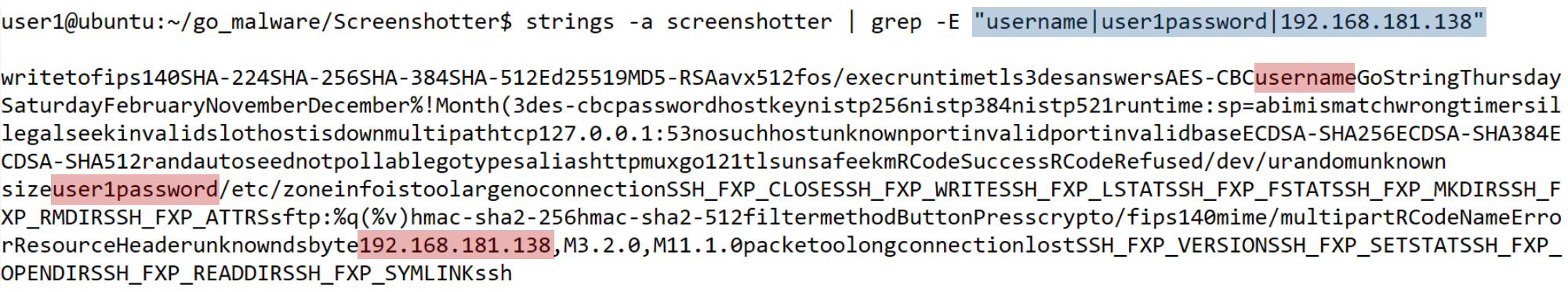}
    \caption{Concatenated output produced by the Linux \texttt{strings} utility}
    \label{go_strings_commands}
  \end{subfigure}

  \begin{subfigure}{0.9\textwidth}
    \centering
    \includegraphics[width=\textwidth]{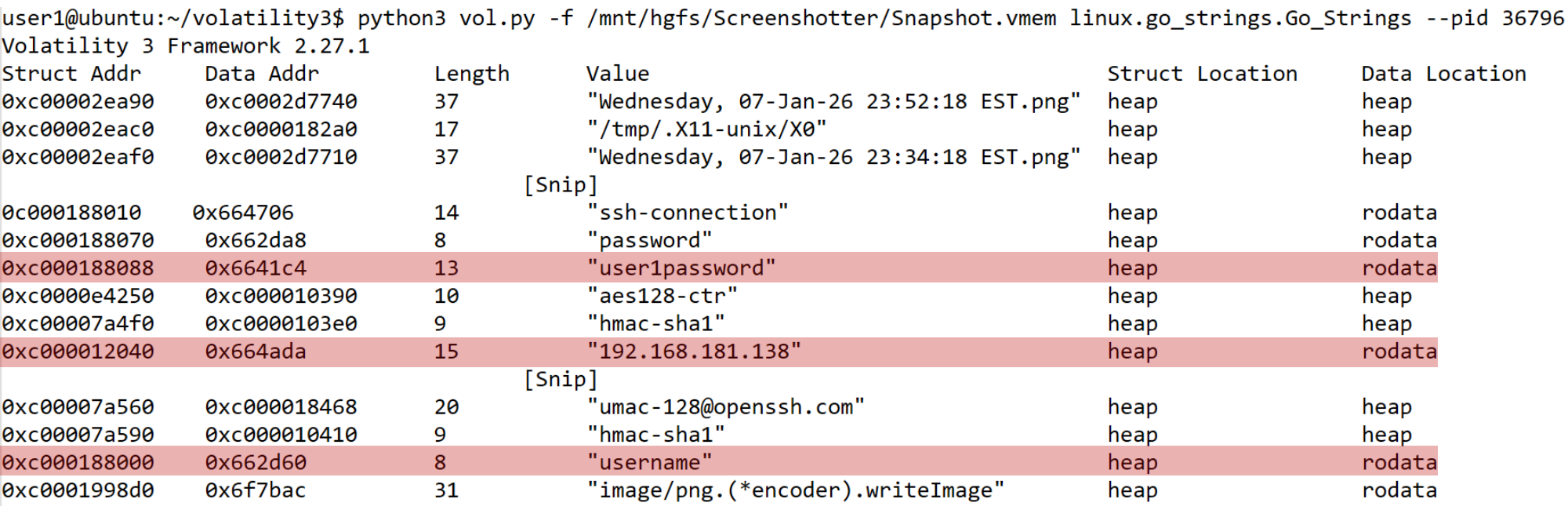}
    \caption{Type-aware strings recovered by the \texttt{go\_strings} plugin}
    \label{go_strings_plugin}
  \end{subfigure}
  \vspace{-0.05in}
  \caption{Comparison of the \texttt{go\_strings} plugin and the Linux \texttt{strings} utility}
  \label{strings}
   \vspace{-0.15in}
\end{figure*}

We demonstrate the advantages of \texttt{go\_strings} over the standard \texttt{strings} utility on the \textit{Screenshotter} sample. In statically linked Go binaries, compile-time string literals are stored as contiguous blobs without null terminators. These blobs combine strings not only from application code, but also from the Go runtime and third-party libraries. As a result, \texttt{strings} produces a single concatenated output stream from which individual artifacts are difficult to recover. As shown in Figure \ref{go_strings_commands}, application-specific artifacts, including the username, password, and remote host used for SFTP-based screenshot exfiltration, are indistinguishable within the output. In contrast, Figure \ref{go_strings_plugin} shows that \texttt{go\_strings} reconstructs Go string headers together with their backing data and memory locations. This enables reliable extraction of sensitive forensic artifacts and identifies whether strings are compile-time constants or generated at runtime.

\vspace{-0.05in}
\subsection{Obscura Ransomware Memory Analysis}
\textit{Obscura} is a Go-based ransomware first observed in August~2025, targeting enterprise environments through domain controller compromise and \texttt{NETLOGON} share distribution \citep{huntress_obscura_2025}. It employs XChaCha20 encryption with Curve25519 key agreement, terminates security software prior to encryption, and deletes volume shadow copies to prevent recovery. We analyzed a memory image from an infected Windows host containing a live Obscura process (PID~1100). Using \texttt{go\_strings}, we reconstructed heap allocation state (112 of 942 spans active) and recovered 778 strings from 550 heap objects, along with 728 additional strings from non-heap regions. 
Table~\ref{obscura_strings} presents runtime artifacts recovered from memory, compared against public threat intelligence reports on Obscura \citep{huntress_obscura_2025, pcrisk_obscura_2025}. Highlighted entries denote artifacts absent from these reports: the 32-byte Curve25519 public key value, which enables decryption if the corresponding private key is recovered; exact Go module dependencies revealing the cryptographic implementation, which link samples compiled from the same build environment; and the Go toolchain (\texttt{GOROOT}) path, which exposes the threat actor's toolchain configuration on a VeraCrypt-encrypted volume. This table also includes previously reported indicators, such as domain-role detection strings (standalone, domain member, BDC, PDC), which provide direct evidence of Obscura's propagation logic, escalating from workstations to domain controllers and subsequently targeting all domain-joined systems when executed on a primary domain controller.
\renewcommand{\arraystretch}{0.95}
\begin{table*}[!htb]
\centering
\caption{Runtime forensic artifacts recovered from Obscura memory using \texttt{go\_strings}. \colorbox{novelartifact}{Highlighted cells} indicate the novel artifacts}

\label{obscura_strings}
\resizebox{0.9\textwidth}{!}{
\begin{tabular}{p{6cm}p{12cm}}
\toprule
\textbf{Artifact} & \textbf{Value} \\
\midrule
Curve25519 Public Key (Base64) & \cellcolor{novelartifact} \texttt{Z7oIV9mbFaMuePnpNBvdcweN+FIFLFVaK1a8TmVMTUs=} \\
Encryption / Key Exchange  Algorithms & \texttt{XChaCha20} / \texttt{Curve25519} \\
\midrule
TOX ID (from the note)  & \texttt{AE55FC0EB1C25A5B081650108F9081E236DECE1CE08D2E185A6F15B9FB48E700210BED374643} \\
Tor Hidden Service (from the note)  & \texttt{http://obscurad3aphckihv7wptdxvdnl5emma6t3vikcf3c5oiiqndq6y6xad.onion} \\
 Victim ID (from the note) & \texttt{148061707215636819} \\
\midrule
Crypto Dependency & \cellcolor{novelartifact} \texttt{golang.org/x/crypto v0.0.0-20201203163018-be400aefbc4c} \\
System Dependency & \cellcolor{novelartifact} \texttt{golang.org/x/sys v0.0.0-20201018230417-eeed37f84f13} \\
 Go Toolchain (GOROOT) Path & \cellcolor{novelartifact} \texttt{/run/media/veracrypt1/Locker Deps/go1.15.linux-amd64/go} \\
Module Path & \texttt{main/windows/locker} \\
\midrule
Domain Role Detection Strings &  \texttt{[+] detect standalone pc.}, \texttt{[+] detect pc in domain. run transfer to dc.}, \texttt{[+] detect BDC. run transfer to PDC.}, \texttt{[+] detect PDC. run transfer to all pc in domain.}\\
\midrule
Runtime Status Messages &  \texttt{[!!!] user not admin. exit [!!!]}, \texttt{[-] failed to delete shadow copies.}, \texttt{[-] failed to run as daemon!}, \texttt{[+] encryption start. you may close this window}, \texttt{Error calling DsRoleGetPrimaryDomainInformation:}\\
\midrule
Security Software & \texttt{WinDefend, MsMpEng, MpCmdRun, CSFalconService, SentinelAgent} \\
Antivirus & \texttt{bdagent, McAfee, SymCorpUI, ccSvcHst, AMService, Emsisoft*} \\

\bottomrule
\end{tabular}
}
\end{table*}

\begin{sloppypar}
Using the \texttt{go\_functions} plugin, we recovered Obscura’s functions together with their source file paths. Similar to compiler-generated PDB paths, file paths are forensically valuable, as they expose characteristics of the build environment that may persist across malware samples \citep{unit42_bookworm}. In Obscura, the recovered build root (\texttt{/run/media/veracrypt1/Backups/Obscura/Locker/windows/}) indicates compilation from a VeraCrypt-mounted volume. Grouping recovered functions by their file paths indicates a modular internal organization. The main module (\texttt{locker/locker.go}) implements file discovery (\texttt{scanDisk}), parallel encryption (\texttt{worker}), ransom note deployment (\texttt{dropNote}), and anti-analysis checks (\texttt{runtimeCheck}); the encryption module (\texttt{api/xchacha20.go}) implements full and partial encryption routines (\texttt{EncryptFull}, \texttt{EncryptPart}); and \texttt{api/api.go} implements Windows API wrappers for privilege checks (\texttt{IsRunAsAdmin}), drive enumeration (\texttt{GetWindowsDrives}), and process termination (\texttt{KillProcess}).
\end{sloppypar}
\subsection{BRICKSTORM Memory Analysis}
\textit{BRICKSTORM} is a Go-based backdoor attributed to the China-nexus threat actor \textit{UNC5221} and has been observed in intrusions targeting VMware vCenter infrastructure, with activity reported through December 2025 \citep{mandiant_brickstorm_espionage_2025,cisa_brickstorm_2025}. Public threat intelligence characterizes its persistence mechanisms, DNS-over-HTTPS (DoH) command-and-control, and post-exploitation capabilities, while providing limited insight into execution-time state and actionable memory-resident artifacts. We analyzed a Linux sample (PID~2004) masquerading as a VMware Java component.

Table~\ref{brickstorm_summary} summarizes quantitative metadata recovered using the \texttt{go\_functions} plugin, including the number of recovered functions and source file paths from both process memory and the cached binary, in addition to the number of recovered types, interfaces, and type methods.
\renewcommand{\arraystretch}{0.9}

\begin{table}[!htb]
\caption{Recovered BRICKSTORM metadata using \texttt{go\_functions}}
\label{brickstorm_summary}
\centering
\resizebox{\columnwidth}{!}{%
\begin{tabular}{ll}
\toprule
\textbf{Attribute} & \textbf{Value} \\
\midrule
Process ID (PID) & 2004 \\
Cached ELF Path & \texttt{/usr/java/jre-vmware/bin/upgrademgr} \\
ELF Architecture & 64-bit, Little Endian \\
ELF Size (Extracted) & 5,844,992 bytes \\
Go Version & \texttt{go1.16.3hijacke} \\
\midrule
Cached Function Names & 5,524 \\
Cached File Paths & 5,524 \\
Recovered Functions & 5,524 \\
Recovered Types & 3,597 \\
Recovered Interfaces & 360 \\
Recovered Type Methods & 1,695 \\
\bottomrule
\end{tabular}
}
\end{table}

Given that the analyzed sample was compiled with Go~1.16, argument recovery relies on parsing \texttt{ArgsPointerMaps} (see Section~\ref{s3-3}). Despite this challenge, Table \ref{brickstorm_functions} shows that \texttt{go\_functions} successfully identifies \textit{BRICKSTORM}'s main functions and execution paths and recovers argument values. We compared all recovered artifacts against the CISA malware analysis report \citep{cisa_brickstorm_2025} and Mandiant's threat intelligence publication \citep{mandiant_brickstorm_espionage_2025}. Highlighted entries in Table~\ref{brickstorm_functions} denote artifacts absent from those reports. These include the configuration file \texttt{/etc/sysconfig/vmp} (CISA documents only the parent directory \texttt{/etc/sysconfig/}), the \texttt{KENSO} and \texttt{ENVLOG} environment variables, the utility function \texttt{wssoft/libs/utils.HasExistProc} (distinct from the documented \texttt{wssoft2} C2 package), raw XOR key material used to recover DoH resolver addresses (Table~\ref{xor_keys_full}), a beacon timer interval of \texttt{0x2260ff9290000} ($\approx$ 7 days), randomized sleep and jitter parameters (0--600 and 0--7200 seconds), the C2 termination keyword \texttt{"exit"}, and exit code \texttt{200}.
\begin{figure*}[!htb]
    \centering
    \includegraphics[width=0.75\textwidth]{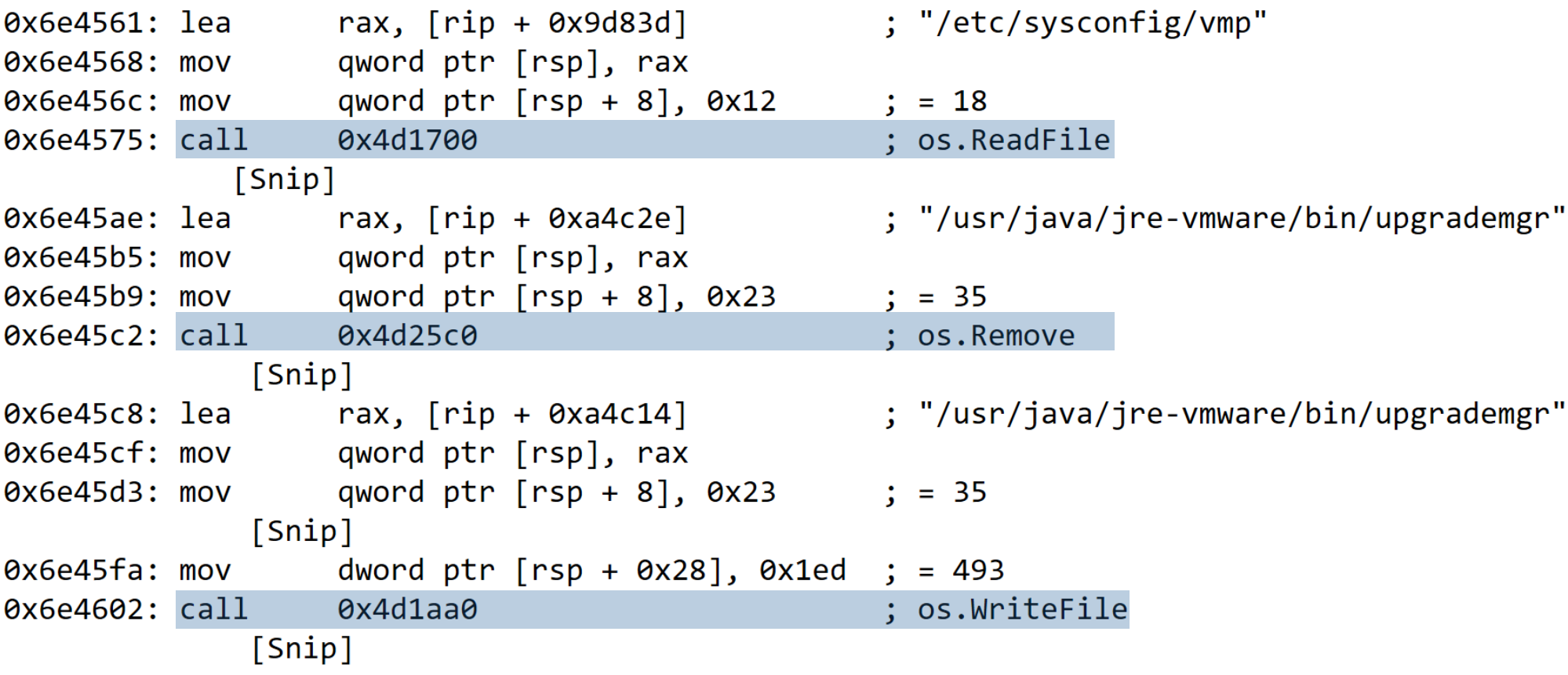}
        \vspace{-0.05in}
   \caption{ABI-Aware Backward Analysis of \texttt{main.copyFile}}
     \vspace{-0.2in}
    \label{copyfile}
\end{figure*}


\begin{figure}
    \centering
    \includegraphics[width=\linewidth]{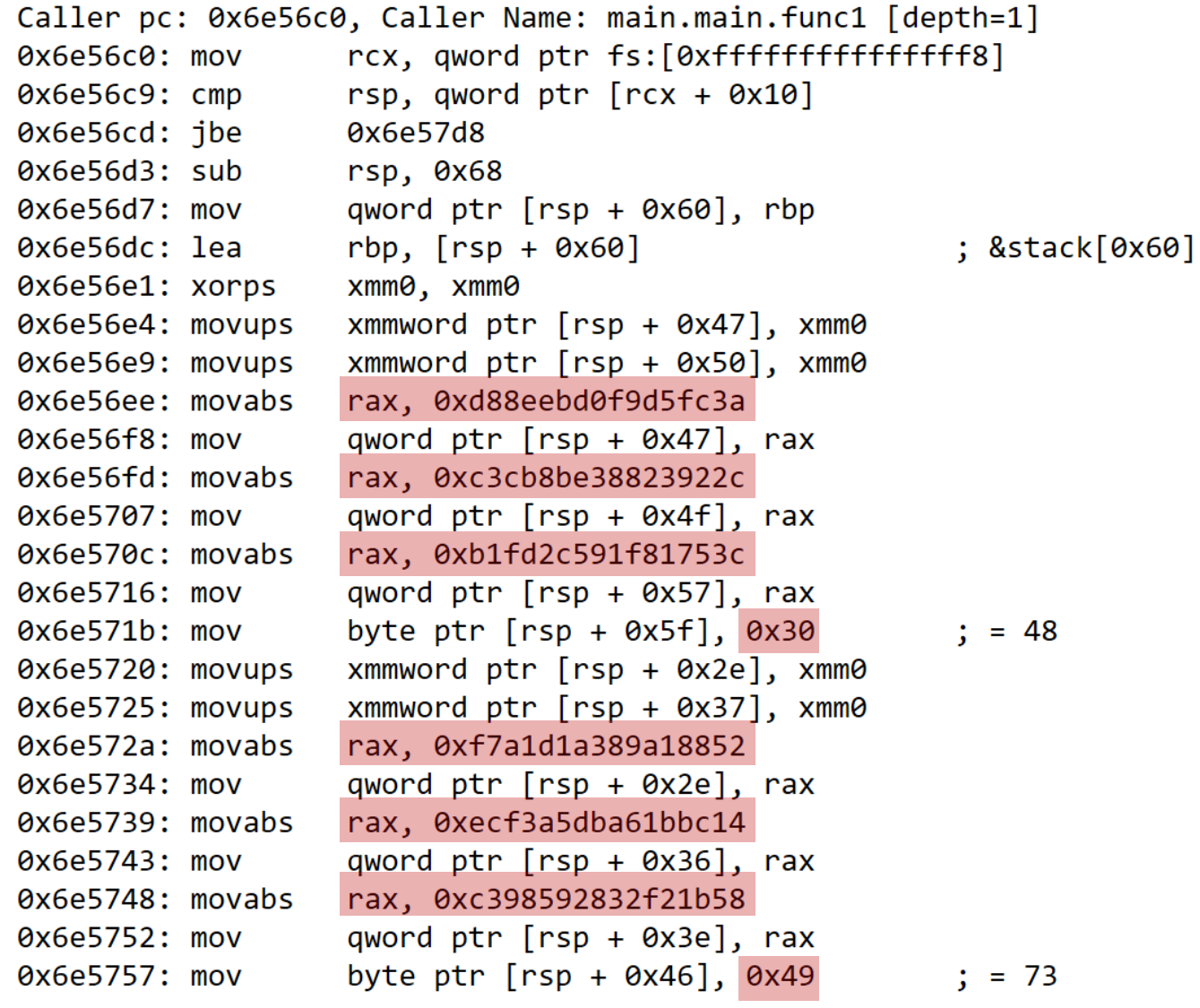}
    \caption{ABI-Aware Backward Analysis of \texttt{main.main.func1}}
    \label{func1}
\end{figure}

\renewcommand{\arraystretch}{0.9}
\begin{table*}[!htb]
\centering
\caption{BRICKSTORM main functions with recovered arguments. \colorbox{novelartifact}{Highlighted cells} indicate novel artifacts discovered by \texttt{go\_functions}}
\label{brickstorm_functions}
\resizebox{0.95\textwidth}{!}{
\begin{tabular}{p{0.7cm}|p{4cm}|p{6cm}|p{6cm}}
\toprule
\textbf{Depth} & \textbf{Function} & \textbf{Recovered Arguments} & \textbf{Forensic Significance} \\
\midrule
\multicolumn{4}{l}{\textit{\textbf{main.main}} (Entry Point) [0x6e4da0]} \\
\midrule
1 & main.startNew & -- & Initialization and persistence \\
\cmidrule{2-4}
& \hspace{0.3cm}$\rightarrow$ os.Getenv &
\texttt{"PATH","TERM","USER"}, \colorbox{novelartifact}{\texttt{"KENSO"}} &
Environment reconnaissance / kill switch \\
& \hspace{0.3cm}$\rightarrow$ main.copyFile & -- & Payload installation \\
\cmidrule{2-4}
2 & \hspace{0.6cm}$\rightarrow$ os.ReadFile & \cellcolor{novelartifact}\texttt{"/etc/sysconfig/vmp"} & Embedded payload retrieval\\
& \hspace{0.6cm}$\rightarrow$ os.Remove & \texttt{"/usr/java/jre-vmware/bin/upgrademgr"} & Pre-installation binary removal \\
& \hspace{0.6cm}$\rightarrow$ os.WriteFile & \texttt{"/usr/java/jre-vmware/bin/upgrademgr", 0755} & \textbf{Trusted-path binary replacement} \\
\cmidrule{2-4}
1 & \hspace{0.3cm}$\rightarrow$ os.Setenv & \texttt{"PATH"}, \texttt{"/usr/java/jre-vmware/bin/"} & PATH hijacking \\
& \hspace{0.3cm}$\rightarrow$ os/exec.Command & \texttt{"upgrademgr"} & Payload execution \\
& \hspace{0.3cm}$\rightarrow$ os.Exit & \texttt{0} & process termination\\
\midrule
\multicolumn{4}{l}{\textit{\textbf{main.selfWatcher}} (Process Watchdog) [0x6e39a0] } \\
\midrule
1 & os.Getenv &  \cellcolor{novelartifact}  \texttt{"ENVLOG"}  & \textbf{Operator-controlled runtime mode} \\

& \cellcolor{novelartifact} wssoft/libs/utils.HasExistProc &  \texttt{"upgrademgr"}  & Process existence verification \\

& os.ReadFile &  \cellcolor{novelartifact} \texttt{"/etc/sysconfig/vmp"} & Payload retrieval for \textbf{self-reinstallation} \\
& os.WriteFile & \texttt{"/usr/java/jre-vmware/bin/upgrademgr", 0755} & self-reinstallation \\
& os.Setenv & \texttt{"PATH"}, \texttt{"/usr/java/jre-vmware/bin/"} & Execution path restoration\\
& os/exec.Command & \texttt{"upgrademgr"} & Watchdog respawn\\

& time.NewTimer &  \cellcolor{novelartifact} \texttt{0x2260ff9290000 ($\approx$ 7 days)}& Long-delay scheduling \\
& math/rand.Int63n &  \cellcolor{novelartifact} \texttt{600 (0-10min)}, \texttt{7200 (0-2hr)} & Randomized sleep and jitter\\
& runtime.selectgo & 3 cases, block=true & Scheduler-mediated multiplexing\\
& os.(*Process).signal & -- & Inter-process signaling\\
\midrule
\multicolumn{4}{l}{\textit{\textbf{main.main} (continued) — C2 Task Loop}} \\
\midrule
1 & main.main.func1--func6 &  \cellcolor{novelartifact} XOR key pairs (Table~\ref{xor_keys_full}) & \textbf{String deobfuscation} \\
& wssoft/core/task.DoTask & \texttt{0x798af8} (task config) & C2 command execution \\

& strings.Index &  \cellcolor{novelartifact} \texttt{"exit"} & C2 termination keyword \\
& time.Sleep & calculated from rand & Evasion delay \\

& os.Exit &  \cellcolor{novelartifact} \texttt{200} & Distinctive termination exit code \\
\bottomrule
\end{tabular}
}
\vspace{-0.05in}
\end{table*}

Analysis of \texttt{main.startNew}  revealed the complete persistence workflow. The malware first checks the \texttt{KENSO} environment variable as a kill-switch; if set to \texttt{"true"}, \textit{BRICKSTORM} removes itself and terminates. Otherwise, it reads an embedded payload from \texttt{/etc/sysconfig/vmp}, deletes any existing binary at \texttt{/usr/java/jre-vmware/bin/upgrademgr}, writes the payload with executable permissions (\texttt{0755}), and modifies the \texttt{PATH} environment variable to prioritize this directory. This masquerades the backdoor as a legitimate VMware Java Runtime component. Notably, these file paths were recovered while analyzing the \texttt{main.copyFile} function (Figure~\ref{copyfile}).  Persistence is maintained via \texttt{main.selfWatcher}, which implements a watchdog that monitors process liveness and reinstalls the payload if necessary. Recovered beacon delay and jitter parameters indicate timing-based evasion, explaining BRICKSTORM’s resistance to sandbox analysis. Prior to C2 communication, six deobfuscation routines (\texttt{main.main.func1}–\texttt{func6}) apply bytewise XOR over stacked key arrays to recover DNS-over-HTTPS resolver endpoints. Figure~\ref{func1} illustrates the XOR key pairs recovered via ABI-aware backward analysis of \texttt{main.main.func1}, while Table~\ref{xor_keys_full} enumerates the complete set of recovered keys and shows their decrypted endpoints.


\renewcommand{\arraystretch}{0.9}
\begin{table*}[!htb]
\centering
\small
\caption{Complete XOR Key Pairs Recovered from \textit{BRICKSTORM} Memory}
\label{xor_keys_full}
\resizebox{0.95\textwidth}{!}{
\begin{tabular}{l|c|l|l|l}
\toprule
\textbf{Function} & \textbf{Len} & \textbf{Key A (hex, little-endian qwords)} & \textbf{Key B (hex, little-endian qwords)}  & \textbf{Decrypted String}\\
\midrule
\texttt{func1} & 25 & \texttt{d88eebd0f9d5fc3a, c3cb8be38823922c, b1fd2c591f81753c, 30} & \texttt{f7a1d1a389a18852, ecf3a5dba61bbc14, c398592832f21b58, 49} &\texttt{https://8.8.8.8/dns-query}\\
\texttt{func2} & 25 & \texttt{1b67d2765502f10c, 54cfcf26fdae3805, 9ac28998a6fea591, 56} & \texttt{3448e80525768564, 7bfbe112d396163d, e8a7fce98b8dcbf5, 2f} &\texttt{https://8.8.4.4/dns-query}\\
\texttt{func3} & 25 & \texttt{74c4b1463d036b04, 26816391cf54809c, 2273df9d12618a88, a7} & \texttt{5beb8b354d771f6c, 09b84da8e16daea5, 5016aaec3f12e4ec, de} &\texttt{https://9.9.9.9/dns-query}\\
\texttt{func4} & 26 & \texttt{da62e026514ae0cb, 6f7aeb38cfedb01f, 74af8230987fe0c5, 139f} & \texttt{f54dda55213e94a3, 5e4bc501e1d49e26, 11daf31deb1184ea, 6aed} &\texttt{https://9.9.9.11/dns-query}\\
\texttt{func5} & 25 & \texttt{c209a43124860499, 461eb55f0c04f616, b95f447c0bef12fe, 15} & \texttt{ed269e4254f270f1, 692f9b6e2235d827, cb3a310d269c7c9a, 6c} &\texttt{https://1.1.1.1/dns-query}\\
\texttt{func6} & 25 & \texttt{8d93c87a64ed5031, 072864390872e841, eff656a7257d0c4e, ea} & \texttt{a2bcf20914992459, 28194a092642c670, 9d9323d6080e622a, 93} &\texttt{https://1.0.0.1/dns-query}\\
\bottomrule
\end{tabular}
}
\vspace{-0.05in}
\end{table*}

\vspace{-0.05in}
\subsection{Pantegana Memory Analysis}
\textit{Pantegana} is a fully-featured, Go-based remote access trojan that supports command execution, file transfer, and system fingerprinting. Originally released as an open-source project, it was later abused by the China-nexus group RedNovember in a global espionage campaign targeting government ministries, defense contractors, and critical infrastructure, after which the repository was removed from GitHub \citep{huntio_pantegana_rat, huntio_ghost_pantegana_2025}. 
To conduct our experiment, we obtained a preserved copy of the original source code, compiled the client and server following the same configuration reported in prior analyses, and executed \textit{Pantegana} within a controlled local network environment. We captured memory  from the "victim" system after executing  several reconnaissance commands from the server. We then used \texttt{go\_functions} to  reconstruct the execution flow that starts at \texttt{main.main} and   proceeds to \texttt{client.RunClient}, which orchestrates the RAT's core functionality by calling \texttt{client.RunFingerprinter} for system reconnaissance, \texttt{client.ClientSetup} for HTTP client initialization, and \texttt{client.RequestCommand} for the C2 polling loop. The latter constructs and dispatches HTTP requests via \texttt{net/http.NewRequestWithContext}, \texttt{net/http.(*Client).Do}, and \texttt{client.Middleware}.

However, \texttt{go\_goroutines}  revealed that these application-level functions were absent from the captured call stacks. While awaiting a C2 response, the goroutine is parked by the Go scheduler with a \texttt{select} state, and  the underlying network file descriptor concurrently enters an \texttt{IO wait} state until data becomes available, unwinding application frames and leaving only blocked standard library frames, such as \texttt{net/http.(*persistConn).roundTrip}.  Despite this, \texttt{go\_goroutines} recovered critical runtime artifacts by inspecting \texttt{http.persistConn} structures referenced from blocked goroutine stack arguments.  It identified 31 goroutines (23 valid), of which 13 were actively involved in HTTP/TLS communication with the C2 server. Table~\ref{pantegana_artifacts} summarizes the recovered artifacts, including the C2 address (\texttt{192.168.20.123:1337}), API endpoints used for system fingerprinting (\texttt{/sysinfo}), command output exfiltration (\texttt{/cmdoutput}), and command polling (\texttt{/getcmd}), a runtime-generated authentication token (\texttt{5112caedfb3a}), server responses, negotiated TLS~1.3 parameters, and X.509 certificate fields extracted from TLS buffers in Goroutine~32. These artifacts are constructed dynamically at runtime and are therefore inaccessible to static analysis.

\renewcommand{\arraystretch}{0.9}
\begin{table*}[!htb]
\caption{Runtime forensic artifacts recovered from Pantegana HTTP/TLS goroutines using \texttt{go\_goroutines}}

\label{pantegana_artifacts}
\centering
\resizebox{\textwidth}{!}{%
\begin{tabular}{c|l|l|p{4cm}|p{7cm}}
\toprule
\textbf{Goroutine ID} & \textbf{Wait Reason} & \textbf{Function (Frame)} & \textbf{Field} & \textbf{Recovered Value} \\
\midrule
\multicolumn{5}{l}{\textit{\textbf{C2 Server Configuration}}} \\
\midrule
25, 29, 31, 33, 39, 57 & select & \texttt{net/http.(*persistConn).writeLoop} & \texttt{cacheKey.\{addr; scheme\}} & \texttt{192.168.20.123:1337}; \texttt{https}\\
\midrule
\multicolumn{5}{l}{\textit{\textbf{HTTP Request Buffers (Runtime-Constructed)}}} \\
\midrule
39 & select & \texttt{net/http.(*persistConn).writeLoop} & \texttt{bw.buf} & \texttt{POST /sysinfo HTTP/1.1\textbackslash r\textbackslash nHost: 192.168.20.123:1337} \\
25, 29, 31, 57 & select & \texttt{net/http.(*persistConn).writeLoop} & \texttt{bw.buf} & \texttt{POST /cmdoutput HTTP/1.1\textbackslash r\textbackslash nHost: 192.168.20.123:1337} \\
33 & select & \texttt{net/http.(*persistConn).writeLoop} & \texttt{bw.buf} & \texttt{GET /getcmd HTTP/1.1\textbackslash r\textbackslash nToken: 5112caedfb3a} \\
\midrule
\multicolumn{5}{l}{\textit{\textbf{HTTP Response Buffers (Network-Received)}}} \\
\midrule
24, 28, 30, 38, 56 & select & \texttt{net/http.(*persistConn).readLoop} & \texttt{br.buf} & \texttt{HTTP/1.1 200 OK\textbackslash r\textbackslash nDate: Sat, 27 Dec 2025 23:49} \\
\midrule
\multicolumn{5}{l}{\textit{\textbf{TLS Session State (Negotiated at Handshake)}}} \\
\midrule
25, 29, 31, 33, 39, 57 & select & \texttt{net/http.(*persistConn).writeLoop} & \texttt{tlsState.Version}, \texttt{tlsState.CipherSuite}, \texttt{tlsState.HandshakeComplete}  & \texttt{772 (TLS 1.3)}\newline \texttt{4865 (TLS\_AES\_128\_GCM\_SHA256)}\newline \texttt{True}  \\
\midrule
\multicolumn{5}{l}{\textit{\textbf{C2 Server Certificate (Network-Received)}}} \\
\midrule
32 & IO wait & \texttt{internal/poll.(*pollDesc).waitRead} & \texttt{[]uint8} (DER) & Subject: C=\texttt{US}, ST=\texttt{Hawaii} \\
\midrule
\multicolumn{5}{l}{\textit{\textbf{Connection Metadata (Kernel State)}}} \\
\midrule
32 & IO wait & \texttt{net.(*netFD).Read} & \texttt{net}; \texttt{family}; \texttt{isConnected}; \texttt{Sysfd} & \texttt{tcp}; \texttt{AF\_INET}; \texttt{True}; \texttt{fd=10}\\
\midrule
\multicolumn{5}{l}{\textit{\textbf{Main Client State}}} \\
\midrule
1 & select & \texttt{net/http.(*persistConn).roundTrip} & -- & Main HTTP client blocked awaiting C2 response \\
\bottomrule
\end{tabular}
}
 \vspace{-0.05in}
\end{table*}

\begin{figure*}[!htb]
    \centering
    \includegraphics[width=\textwidth]{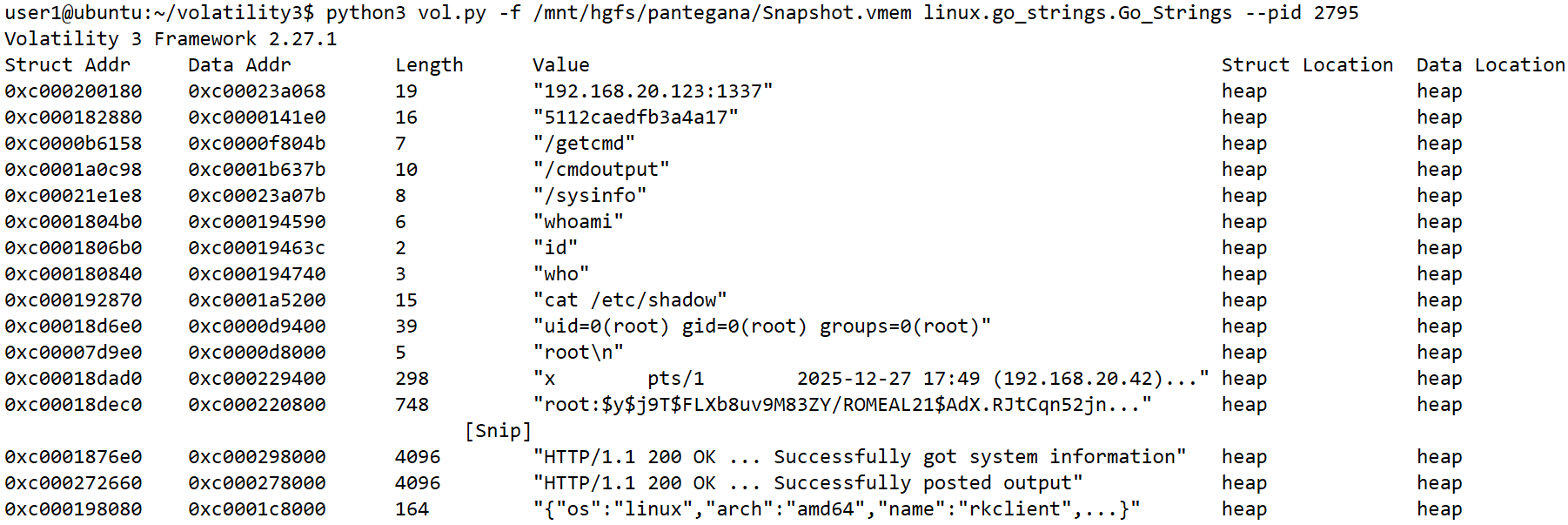}
   \caption{Runtime forensic artifacts recovered from Pantegana memory using \texttt{go\_strings}}
    \label{go_strings_pantegana}
   \vspace{-0.1in}
\end{figure*}

As shown in Figure~\ref{go_strings_pantegana}, \texttt{go\_strings} recovered further heap-resident artifacts, including the C2 IP address, API endpoints used for system fingerprinting (\texttt{/sysinfo}), all of our reconnaissance commands (\textit{whoami}, \textit{id}, \textit{who}, \textit{cat /etc/shadow}), and their complete output. In total, \texttt{go\_strings} recovered all the dynamic aspects of the malware, including  C2 configuration, attacker-issued commands, exfiltrated outputs, and HTTP requests and responses.

\section{Limitations and Future Work}
\label{s5}
Our framework currently targets x86-64 architecture, as it covers the majority of environments where Go malware is observed. Supporting ARM64 would require adapting the ABI-aware backward analysis to different calling conventions and register sets. To this end, large language models present a promising direction for generalizing across architectures.
Moreover, the current analysis focuses on argument recovery at call sites and within goroutine frames. Thus, extending dataflow tracking to local variables and return values would enable richer semantic analysis, including taint propagation and more complete execution reconstruction.
\section{Conclusion}
\label{s6}
This paper presented our efforts in analyzing Go binaries in memory. Our proposed framework parses Go's embedded structures to recover function metadata, type descriptors, and interface tables, considering stripped and obfuscated binaries. Leveraging this metadata, it identifies heap objects, extracts dynamically allocated and static strings, and classifies functions by origin. Through ABI-aware backward analysis, it further reconstructs execution paths and argument values from call sites. Goroutine analysis extends these capabilities by capturing active functions and their runtime argument values.
We implemented these capabilities as Volatility 3 plugins and evaluated them against real-world malware, such as \textit{BRICKSTORM}, \textit{Obscura}, and \textit{Pantegana}. The results demonstrated recovery of critical forensic artifacts, such as C2 configurations, cryptographic keys, persistence mechanisms, and threat actor communication channels, including indicators and behaviours not reported in published threat intelligence.
\vspace{-0.1in}
\bibliographystyle{elsarticle-harv}
\bibliography{main}
\end{document}